\documentclass{emulateapj}

\usepackage{amssymb}
\usepackage{amsmath}
\usepackage{natbib}
\usepackage{txfonts}
\usepackage[colorlinks=true,citecolor=blue,linkcolor=blue]{hyperref}
\usepackage{microtype}
\usepackage{color}
\usepackage{graphicx}
\usepackage{epstopdf}
\def\mnras{MNRAS}
\def\apj{ApJ}
\def\jcap{JCAP}
\def\apjl{ApJL}
\def\apjs{ApJS}

\def\aap{A\&A}

\def\inpe{1}

\shorttitle{Gravitational waves from pulsars and their braking indices: the role of a time dependent magnetic ellipticity}

\shortauthors{de Araujo, Coelho and Costa}

\begin{document}

\title{Gravitational waves from pulsars and their braking indices: the role of a time dependent magnetic ellipticity}

\author{Jos\'e C. N. de Araujo\altaffilmark{\inpe}, Jaziel G. Coelho\altaffilmark{\inpe} and C\'esar A. Costa\altaffilmark{\inpe} }

\altaffiltext{\inpe}{Divis\~ao de Astrof\'isica,        Instituto Nacional de Pesquisas Espaciais, Avenida dos Astronautas 1758,                     12227--010 S\~ao Jos\'e dos Campos,            SP, Brazil}

\begin{abstract}
We study the role of time dependent magnetic ellipticities ($\epsilon_{B}$) on the calculation of the braking index of pulsars. Moreover, we study the consequences of such a $\epsilon_{B}$ on the amplitude of gravitational waves (GWs) generated by pulsars with measured braking indices. We show that, since the ellipticity generated by the magnetic dipole is extremely small, the corresponding amplitude of GWs is much smaller than the amplitude obtained via the spindown limit.
\end{abstract}


\altaffiltext{}{jcarlos.dearaujo@inpe.br; jaziel.coelho@inpe.br; cesar.costa@inpe.br}

\maketitle

\section{Introduction}
\label{int}
Recently, gravitational waves (GWs) have been directly  detected~\citep{2016PhRvL.116f1102A} for the first time. This signal, named GW150914, has been identified as coming from the final fraction of a second of a coalescence of two black holes (BHs), which resulted in a spinning remnant BH. Such an event, though predicted, has not been observed yet via any other means. Even more recently, a second signal, named GW151226, has been detected~\citep{PhysRevLett.116.241103}, which has also been identified as coming from the coalescence of two BHs. This second event reinforces that we are witnessing the dawn of a new field in astronomy, namely, the field of the GW astronomy.

Rapidly rotating neutron stars are also promising candidates for GW signals, which could be detected by Advanced LIGO (aLIGO) and Advanced Virgo (AdVirgo) in the near future. It is well known that these sources might generate continuous GWs whether or not they are perfectly symmetric around their rotation axes. 

In the context of pulsars, the so-called braking index $n$, which is a quantity closely related to pulsar's spindown, can provide information about pulsars' energy loss mechanisms. Such mechanisms can include, among others, GW emission. 
Since pulsars can also spindown through GW emission associated with asymmetric deformations~\citep[see, e.g.,][]{1969ApJ...158L..71F,1969ApJ...157.1395O}, it is appropriate to take into account this mechanism in a model that aims to explain the measured braking indices. Recently,~\cite{2016ApJ...819L..16A} showed that PSR J1640--4631 is the first pulsar with a braking index greater than three, namely $n=3.15\pm 0.03$. 
Until very recently, only eight of the $\sim 2400$ known pulsars have braking indices accurately measured. All these braking indices are remarkably smaller than the canonical value $(n = 3)$, which is expected for pure magneto-dipole radiation model \citep[see, e.g.,][]{1993MNRAS.265.1003L,1996Natur.381..497L,2007ApSS.308..317L,2011ApJ...741L..13E,2011MNRAS.411.1917W,2012MNRAS.424.2213R, 2015ApJ...810...67A}. 
Several interpretations for the observed braking indices have been put forward, like the ones that propose either accretion of fall-back material via a circumstellar disk \citep{2016MNRAS.455L..87C}, the so-called quantum vacuum friction (QVF) effect~\citep{2016ApJ...823...97C}, relativistic particle winds \citep{2001ApJ...561L..85X,2003A&A...409..641W}, or modified canonical models to explain the observed braking index ranges \citep[see e.g.,][and references therein for further models]{1997ApJ...488..409A,2016ApJ...823...34E}.
Another possibility is that the magnetic moment of the star changes in time, through either a change in the surface
field strength or the angle between the magnetic and
spin axes~\citep[see, e.g.,][and references therein]{1995ApJ...440L..77M,1996A&A...308..507P,1998ApJ...508..838L,2000ASSL..254...95E}.
Following this line, it was advanced in~\citep{2016arXiv160305975D,2016EPJC...76..481D} that the appropriate combination of gravitational and electromagnetic contributions on the spindown could explain the measured braking indices. Because of that, we model the braking indices of these pulsars taking into account the spindown due to magnetic dipole and GW brakes, besides considering either the surface magnetic dipole and the angle between the magnetic and rotation axis being time dependent. 

Based on the above discussion, the aim of the present paper is to extend the analysis of~\cite{2016arXiv160305975D,2016EPJC...76..481D} focusing mainly on the role of a time-dependent magnetic ellipticity ($\epsilon_{B}$) on the calculation of the pulsars' braking indices. Here, we show that it is possible to obtain useful equations to calculate the so called {\it efficiency} $\eta$ (or the fraction of deceleration related to GW emission),  $\epsilon_{B}$ and the amplitude of the GWs.

It is worth mentioning, that a time dependent $\epsilon_{B}$ stems naturally from the fact that such a quantity depends on the strength of magnetic field and on the angle between the magnetic and rotation axes, which, as already mentioned, depend on time in our approach. 

The paper is organized as follows. Section \ref{sec:2} is devoted to a brief investigation of the deformation of a pulsar by its magnetic field. In Section~\ref{sec:3} we revisit the associated energy loss focusing mainly on the energy balance, when both gravitational and classic dipole radiations are responsible for the pulsar's spindown. Also, we elaborate upon the evolution of other pulsars' characteristic parameters (i.e., the mean surface magnetic field $B_0$ and the magnetic dipole direction $\phi$), and we include now the role of a time dependent magnetic ellipticity. In Section~\ref{sec:4} we consider the gravitational radiation emitted by a rotating star, distorted by its internal magnetic field. Finally, in Section~\ref{sec:5}, we summarize the main conclusions and remarks. In this paper, we work with Gaussian units.

\begin{table*}
\caption{Periods ($P$) and Their First Derivatives ($\dot P$) for Pulsars with Known Braking Indices ($n$).}
\label{ta1}
\begin{tabular*}{\textwidth}{@{\extracolsep{\fill}}lccccccr@{}}
\hline
Pulsar & $P$~(s) &$\dot{P}~(10^{-13}$~s/s) &$n$ & References & $\epsilon$ & $\eta$   \\ 
\hline
PSR J1734-3333      &1.17  &22.8 &$0.9\pm0.2$       &~\cite{2011ApJ...741L..13E} & $1.2\times 10^{-7(-5)} $  & $1.1\times 10^{-13(-9)} $ \\
PSR B0833-45 (Vela) &0.089 &1.25 & $1.4\pm0.2$      &~\cite{1996Natur.381..497L} & $4.9\times 10^{-10(-8)}$  & $8.3\times 10^{-14(-10)}$ \\
PSR J1833-1034      &0.062 &2.02 &$1.8569\pm0.0006$ &~\cite{2012MNRAS.424.2213R} & $5.5\times 10^{-10(-8)}$  & $1.9\times 10^{-13(-9)} $ \\
PSR J0540-6919      &0.050 &4.79 &$2.140\pm0.009$   &~\cite{2007ApSS.308..317L}  & $1.1\times 10^{-9(-7)} $  & $5.7\times 10^{-13(-9)} $ \\ 
PSR J1846-0258      &0.324 &71   &$2.19\pm0.03$     &~\cite{2015ApJ...810...67A} & $1.0\times 10^{-7(-5)} $  & $1.3\times 10^{-12(-8)} $ \\
PSR B0531+21 (Crab) &0.033 &4.21 &$2.51\pm0.01$     &~\cite{1993MNRAS.265.1003L} & $6.1\times 10^{-10(-8)}$  & $7.5\times 10^{-13(-9)} $ \\
PSR J1119-6127      &0.408 &40.2 &$2.684\pm0.002$   &~\cite{2011MNRAS.411.1917W} & $7.2\times 10^{-8(-6)} $  & $5.8\times 10^{-13(-9)} $ \\
PSR J1513-5908      &0.151 &15.3 &$2.839\pm0.001$   &~\cite{2007ApSS.308..317L}  & $1.0\times 10^{-8(-6)} $  & $6.0\times 10^{-13(-9)} $ \\ 
PSR J1640-4631      &0.207 &9.72 &$3.15\pm0.03$     &~\cite{2016ApJ...819L..16A} & $8.9\times 10^{-9(-7)} $  & $2.8\times 10^{-13(-9)} $ \\
\hline
\end{tabular*}
\\Note.  Also shown are $\epsilon$ and $\eta$ for $\kappa = 10\; (1000)$.
\end{table*}

\section{Ellipticity of magnetized stars}
\label{sec:2}
This section deals with the pulsars' deformations induced by a strong magnetic field, relying mainly on the seminal works of~\cite{1953ApJ...118..116C}, ~\cite{1996A&A...312..675B}, and ~\cite{2000A&A...356..234K}.
In this regard, GW emission from magnetic distorted stars was duly discussed by~\cite{2006A&A...447....1R} and ~\cite{2001A&A...367..525P}. The aforementioned distortion is supposed to be symmetric around some axis inclined with respect to the rotation axis.
In order to investigate the effect arising from magnetic stress on the equilibrium of stars, let us introduce the fiducial equatorial ellipticity, defined as~\citep[see, e.g.,][]{2001thas.book.....P,1983bhwd.book.....S,2014ApJ...785..119A}
\begin{equation}
\epsilon=\frac{I_{xx}-I_{yy}}{I_{zz}},
\end{equation}
where $I_{xx}$, $I_{yy}$, and $I_{zz}$ are the moment of inertia with respect to the rotation axis $z$, and along directions perpendicular to it.
It was shown by  \cite{1953ApJ...118..116C} that the figure of equilibrium of an incompressible fluid sphere with an internal uniform magnetic field that matches an external dipole field, is not represented by a sphere. The star becomes oblate by contracting along the axis of symmetry, namely along the direction of the magnetic field~\citep[see, e.g.,][]{2014ApJ...794...86C}. Thus, we consider that the fluid sphere is deformed in such a way that the equation of the ellipticity arising from the magnetic field is given by~\citep{1996A&A...312..675B,2000A&A...356..234K,2006A&A...447....1R}
\begin{equation}
\epsilon_B = \kappa\frac{B_0^2 R^4}{G M^2}\sin^2\phi, \label{eq:epsilonB}
\end{equation}
where $B_0$ is the dipole magnetic field, $R$ and $M$ are, respectively, the radius and mass of the star, $\phi$ is the angle between the spin and magnetic dipoles axes, while the factor $\kappa$ is the distortion parameter, which depends on both the star equation of state (EoS) and the magnetic field configuration. 

In the next sections, we consider the role of a time dependent $\epsilon_{B}$ on the calculation of pulsars braking index; and after that we evaluate $\epsilon_{B}$ and $\eta$ as well as their consequences on the calculations of the GW amplitudes generated.

\section{Modeling pulsars' braking indices: the role of a time dependent magnetic dipole ellipticity }\label{sec:3}

In this section, we consider the role of the time dependent $\epsilon_{B}$, discussed in the previous section, in the modeling of the pulsars' braking indices.

To proceed, we follow one of our previous papers closely~\citep{2016arXiv160305975D,2016EPJC...76..481D} in which a detailed derivation of the braking index taking, taking into account the magnetic dipole brake, as well as a GW brake.   
As it is well known, and already discussed in the aforementioned previous works, if the pulsar magnetic dipole moment is misaligned with respect to its spin axis by an angle $\phi$, the energy per second emitted by the rotating magnetic dipole is given by \citep[see, e.g.,][]{1975ctf..book.....L, 2001thas.book.....P},
\begin{equation}
\dot{E}_{\rm d}= -\frac{16\pi^4}{3}\frac{B_0^2 R^6\sin^2\phi}{c^3}f_{\rm rot}^4,  \label{Ed}
\end{equation}
where $R$ is the radius of the star, $f_{\rm rot}$ is the rotational frequency, and $c$ is the speed of light.

On the other hand, a spheroidal body with moment of inertia, $I$, and equatorial ellipticity, $\epsilon$, emits GWs. In this case, the energy loss via GW emission reads~\citep[see, e.g.,][]{1983bhwd.book.....S}
\begin{equation}
\dot{E}_{\rm GW} = -\frac{2048\pi^6}{5}\frac{G}{c^5}I^2\epsilon^2 f_{\rm rot}^6. \label{EGW}
\end{equation}

Now, we consider that the total energy of the star is provided by its rotational energy, $E_{\rm rot} = 2  \pi^2If_{\rm rot}^2$, and any change on it is attributed to both $\dot{E}_{\rm d}$ and $\dot{E}_{\rm GW}$. Therefore, the energy balance reads
\begin{equation}
\dot{E}_{\rm rot}\equiv \dot{E}_{\rm GW} +\dot{E}_{\rm d} \label{Erotdef},
\end{equation}
consequently, it follows immediately that 
\begin{equation}
\dot{f}_{\rm rot} = - \frac{512\pi^4}{5} \frac{G}{c^5}I\epsilon^2f^5_{\rm rot} - \frac{4\pi^2}{3}\frac{B_0^2R^6\sin^2\phi}{Ic^3}f^3_{\rm rot}. \label{domeg}
\end{equation}

Now, we can obtain the equation for the braking index $n$  whose definition reads
\begin{equation}
n = \frac{f_{\rm rot}\,{\ddot f}_{\rm rot}}{\dot{f}^2_{\rm rot}}\label{n}.
\end{equation}

Recall that a pure magnetic brake, in which a dipole magnetic configuration is adopted, leads to $n = 3$, whereas a pure GW brake leads to $ n = 5$~\citep{1983bhwd.book.....S}. From the observational point of view, the literature shows that almost all pulsars with measured braking indices have $ n < 3$ (see Table \ref{ta1}). However, there is one exception: PSR J1640--4631 presents a braking index $ n \simeq 3.15$. Therefore, neither a pure GW brake nor a pure magnetic dipole brake are supported by the observations.

We have recently shown that the braking index of PSR J1640--4631 can be accounted for a combination of GWs and magnetic dipole brake~\citep[see][]{2016arXiv160305975D}. As it is well known, a possible way to explain brake indices $ n < 3$ considers that the magnetic field and/or the angle between the rotation and magnetic axis are time dependent. Since we consider in the present paper that the ellipticity has a magnetic origin, a consequence thereof is that this quantity is time dependent too.

To proceed, by substituting equation \ref{domeg} and its first derivative into equation \ref{n}, the braking index reads
\begin{equation}
n=n_0+ \frac{f_{\rm rot}}{{\dot f}_{\rm rot}}\left(n_0-1\right)\left[\frac{\dot B_0}{B_0}+\dot{\phi}\cot{\phi}  \right], \label{nBphi}
\end{equation}
with
\begin{equation}
n_0 = 3 + \frac{2}{1+\frac{5}{384}\frac{c^2B_0^2R^6\sin^2\phi}{G\pi^2 I^2\epsilon^2f_{\rm rot}^2}},
\end{equation}
where we consider that $\dot{B_0}$ and $\dot{\phi}$ are not null and consequently $\epsilon_{B}$ depends on time. Notice that if one considered that $\epsilon_{B}$ did not depend on time the term $(n_0 -1)$ would be substituted by $(5-n_0)$  \citep[see][]{2016EPJC}.

The above equation for $n$ can be rewritten in terms of the efficiency of the generation of GWs. This quantity stems naturally from the following reasoning. Equation \ref{domeg} can be interpreted as follows: the term on the left side stands for the resulting deceleration (spindown) due to magnetic dipole and GW brakes, the terms on the right side denote the independent contributions of these decelerating processes, respectively. Then, equation \ref{domeg} can be rewritten in the following form
\begin{equation}
\dot{f}_{\rm rot} = \dot{f}_{\rm GW} + \dot{f}_{\rm d}. 
\end{equation}
As a realization of our analyses, let us define now the fraction of deceleration related to GW emission, namely
\begin{equation}
\eta \equiv \frac{\dot{f}_{\rm GW}}{\dot{f}_{\rm rot}}, \label{eta0}
\end{equation}
which, by replacing the appropriate quantities, reads
\begin{equation}
\eta = \frac{1}{1+\frac{5}{384}\frac{c^2B_0^2R^6\sin^2\phi}{G\pi^2 I^2\epsilon^2f_{\rm rot}^2}}. \label{eta}
\end{equation}
From equation \ref{eta0}, it follows immediately that
\begin{equation}
\dot{E}_{\rm GW} = \eta \dot{E}_{\rm rot},
\end{equation}
thus $\eta$ can be interpreted as the efficiency of GW generation.  

\begin{figure}
\includegraphics[width=\linewidth,clip]{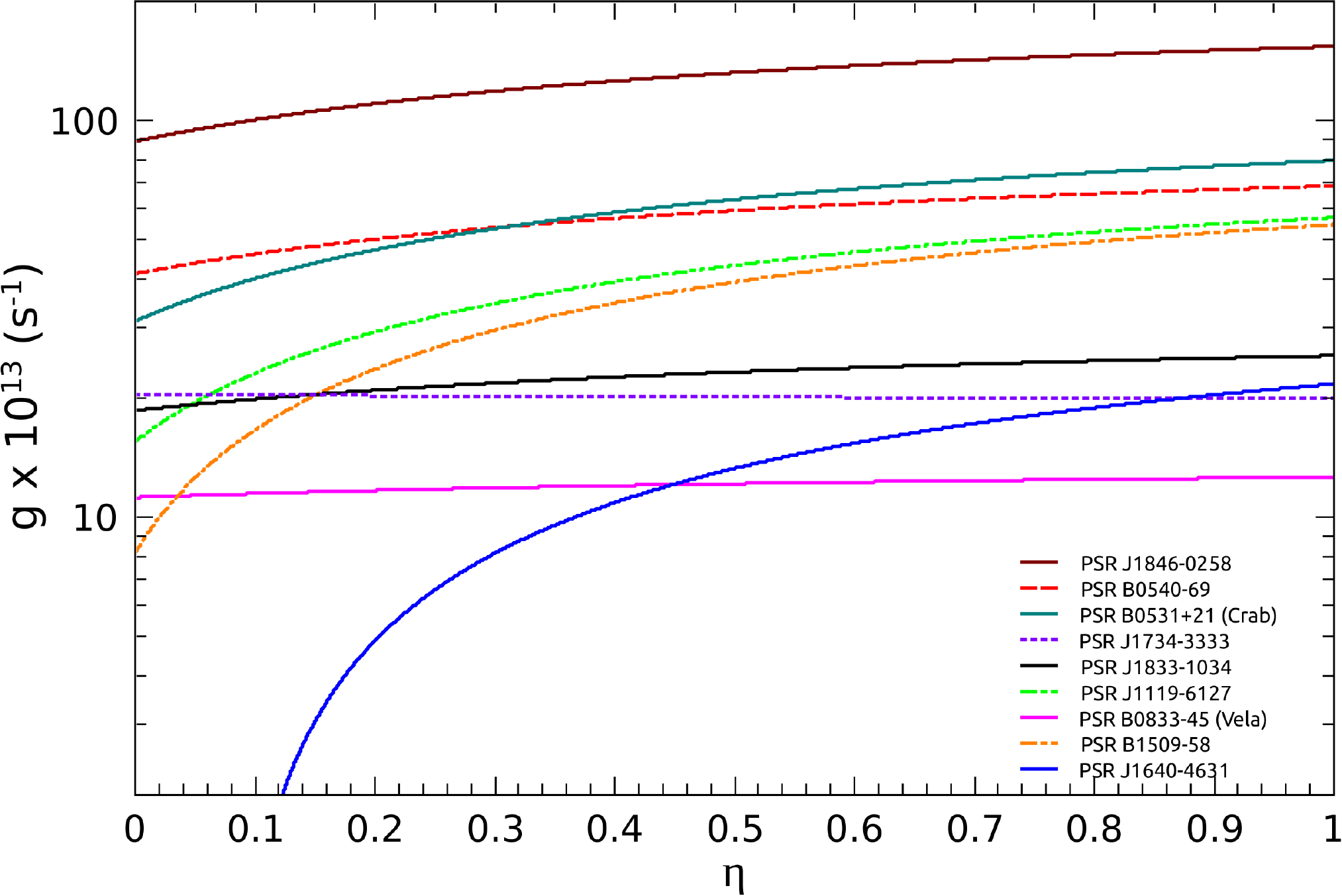}
\caption{The term in brackets ($g$) as a function of $\eta$.}\label{fig1}
\end{figure}

Notice that, from equations \ref{nBphi} and \ref{eta}, it follows immediately that $n_0 = 3 + 2\eta$. Consequently, we finally obtain an equation that relates the braking index to the efficiency of generation of GWs, namely
\begin{equation}
n=3+2\eta-2\frac{P}{\dot P}\left(1+\eta\right)\left[\frac{\dot B_0}{B_0}+\dot{\phi}\cot{\phi}  \right], \label{neta}
\end{equation}
conveniently written in terms of the rotational period ($P = 1/f_{\rm rot}$) and its first derivative ($\dot{P}$), in order to be directly applied to the data of Table \ref{ta1}. Notice that the above equation shows that, in principle, it is possible to obtain $ n < 3$ if the appropriate combination of $\dot{B_0}$ and $\dot{\phi}$ turns the term in brackets positive.
In order to proceed, it is interesting to calculate the term in brackets as a function of $\eta$ for the pulsars of Table \ref{ta1}. For the sake of simplicity, the term in brackets is rewritten as follows,
\begin{equation}
g = g(B_0,\dot{B}_0,\phi,\dot{\phi})\equiv \left[\frac{\dot B_0}{B_0}+\dot{\phi}\cot{\phi}  \right]. \label{gb}
\end{equation}
Thus, the term in brackets as a function of $\eta$ for a given pulsar reads
\begin{equation}
g = -\frac{(n-3-2\eta)}{2(1+\eta)}\frac{\dot{P}}{P}.
\end{equation}
In Figure \ref{fig1}, we present the term in brackets ($g$) as a function of $\eta$.
This figure shows that it is, in principle, possible, as already mentioned, to find a suitable combination of $\dot{B_0}$ and $\dot{\phi}$ in order to have $ n < 3$ and GWs be generated.  

It is believed that magnetic fields should decay in pulsars, usually due to the Ohmic decay, Hall drift, and ambipolar diffusion~\citep{1988MNRAS.233..875J,1992ApJ...395..250G} on timescales of the order of $(10^{6}-10^{7})$ years~\citep[see, e.g.,][and references therein]{1992ApJ...395..250G,2015MNRAS.453..671G}. Nevertheless, there are also suggestions that the timescales for $B_0$ decay could actually be smaller, of the order of $10^5$ years~\citep{2014MNRAS.444.1066I,2015AN....336..831I}. However, timescales of the order of $\sim10^8$ yrs have been suggested, based on population synthesis techniques~\citep[see][]{1997ApJ...489..928M,2000A&A...359..242R,2001A&A...374..182R}. Also, numerical simulations by \citet{1992A&A...254..198B},~\citet{1997A&A...322..477H}  and~\citet{1997ApJ...489..928M} suggest that
the observed properties of the pulsar population are consistent with decay times longer than the pulsar lifetime.

Thus, bearing in mind that $ B_0\sim 10^{12}-10^{13}\,\rm{G}$, let us assume $\dot{B}_0<0$ and $|\dot{B}_0| \sim 10^{-2}-10^{-1}\,\rm{~G/s}$~\citep[see, e.g.,][]{2016ApJ...823...97C}. Since the Crab pulsar has an observationally inferred $\dot\phi\simeq 3\times 10^{-12}\,\rm{rad/s}$~\citep{2013Sci...342..598L, 2015MNRAS.446..857L,2015MNRAS.454.3674Y,2016ApJ...823...97C}, let us consider the implications of these parameters. For instance, consider the representative angle $\phi = \pi/4$  and $\dot{B}_0= -0.05\,\rm{G/s}$, from which we obtain $g\simeq 3\times 10^{-12}\,\rm{s}^{-1}$.

Notice that PSR J1640--4631 can also have its braking index $n = 3.15$ consistently explained. In our previous paper~\citep[see][]{2016arXiv160305975D}, $\eta = 0.075$, and in the present model we can have $0 \leq \eta \leq 1$, depending on the values of $\dot{B_0}$ and $\dot{\phi}$.

\section{Calculating $\eta$, $\epsilon_{B}$ and the amplitude of GWs}
\label{sec:4}
In this section, we show that it is possible to obtain useful equations to calculate $\eta$, $\epsilon_{B}$ and the amplitude of the GWs.

Recall that one usually finds in the literature the following equation
\begin{equation}
h^2 = \frac{5}{2}\frac{G}{c^3}\frac{I}{r^2}\frac{\mid\dot{f}_{\rm rot}\mid}{f_{\rm rot}},
\end{equation}
\citep[see, e.g.,][]{2014ApJ...785..119A}, where one is considering that the whole contribution to $\dot{f}_{\rm rot}$ comes from the GW emission, which means that one is implicitly assuming that $ n = 5$. This equation must be modified to take into account that $n < 5$. 

From equation \ref{eta0} we can write
\begin{equation}
\dot{\bar{f}}_{\rm rot} = \eta \dot{f}_{\rm rot}, 
\end{equation}
where $\dot{\bar{f}}_{\rm rot}$ can be interpreted as the part of $\dot{f}_{\rm rot}$ related to the GW emission brake. Thus, the GW amplitude is now given by
\begin{equation}
\bar{h}^2 = \frac{5}{2}\frac{G}{c^3}\frac{I}{r^2}\frac{\mid\dot{\bar{f}}_{rot}\mid}{f_{rot}} =  \frac{5}{2}\frac{G}{c^3}\frac{I}{r^2}\frac{\mid\dot{f}_{\rm rot}\mid}{f_{\rm rot}} \, \eta . \label{heta}
\end{equation}
On the other hand, recall that the amplitude of GWs can also be written as follows
\begin{equation}
h = \frac{16\pi^2G}{c^4} \frac{I\epsilon f_{\rm rot}^2}{r},
\end{equation}
\citep[see, e.g.,][]{1983bhwd.book.....S}, which with the use of equation \ref{heta} yields an equation for $\epsilon$ in terms of $P$, $\dot P$ (observable quantities), $\eta$ and $I$, namely
\begin{equation}
\epsilon = \sqrt{\frac{5}{512\pi^4} \frac{c^5}{G}\frac{\dot{P}P^3}{I}\eta}. \label{epet}
\end{equation}

Recall also that, for a pure magnetic brake, one can readily write that 
\begin{equation}
\bar{B}_0\sin^2\phi = \frac{3 I c^3}{4 \pi^2 R^6} P \dot{P}
\end{equation}
where $\bar{B}_0$ would be the magnetic field whether the break is magnetic only. In the case in which there is also a GW brake contribution, one has $B_0 < \bar{B}_0$. Notice that the appropriate combination of equations \ref{eta} and \ref{epet} provides the following equation for the efficiency $\eta$ 
\begin{equation}
\eta = 1 - \left(\frac{B_0}{\bar{B}_0} \right)^2,
\end{equation}
which is obviously smaller than one, as it should be. Substituting this last equation into equation \ref{eq:epsilonB} one immediately obtains that
\begin{equation}
\epsilon_B = \frac{3Ic^3}{4\pi^2GM^2R^2}P\dot{P} \left(1 - \eta\right) \kappa. \label{eek}
\end{equation}
Finally, by substituting this last equation into equation \ref{epet}, one immediately obtains that 
\begin{equation}
\eta = \frac{288}{5}\frac{I^3c}{GM^4R^4}\frac{\dot{P}}{P}\left( 1-\eta \right)^2 \kappa^2. \label{ek}
\end{equation}

Notice that equations \ref{eek} and \ref{ek} allow us to obtain $\epsilon_{B}$ and $\eta$ in terms of $M$, $R$, $I$, $P$, and $\dot{P}$ for any given value of $\kappa$. Since in practice $\eta \ll 1$, one can readily obtain the following useful equations 
\begin{equation}
\epsilon_B \simeq \frac{3Ic^3}{4\pi^2GM^2R^2}P\dot{P}\kappa
\label{eek2}
\end{equation}
and 
\begin{equation}
\eta \simeq \frac{288}{5}\frac{I^3c}{GM^4R^4}\frac{\dot{P}}{P} \kappa^2. \label{ek2}
\end{equation}

Now, we are ready to calculate $\epsilon_{B}$ and $\eta$ for the pulsars of Table \ref{ta1}. To do so, we can adopt fiducial values for $M$, $R$, and $I$. Regarding the distortion parameter $\kappa$, as already mentioned, it depends on the EoS and the magnetic field configuration. In particular, we chose $\kappa = 10$ and $\kappa = 1000$, which have the same orders of magnitude of the values considered by, for example, \cite{2006A&A...447....1R}. It is worth noting that the higher value of $\kappa$ adopted is probably unrealistic~\citep[see, e.g.,][for a brief discussion]{2006A&A...447....1R}. In the last two columns of Table \ref{ta1} we present the result of these calculations. Notice that, even considering an extremely optimistic case, the value of the ellipticity is at best $\epsilon_{B} \sim 10^{-5}$ (for PSR J1846--0258) and the corresponding efficiency $\eta \sim 10^{-8}$. Thus, the GW amplitude in this case would be four orders of magnitude lower than the amplitude obtained by assuming the spindown limit ($\eta =1$). 
Since the predicted GW amplitudes are extremely small
for all pulsars of Table \ref{ta1}, even advanced detectors such as aLIGO and AdVirgo, and the planned Einstein Telescope (ET) would not be able to detect these pulsars, whether or not the ellipticity is of magnetic dipolar origin.
For example, if we consider again \mbox{PSR J1846--0258}, in its best scenario, and recalling that the sensitivity depends on the squared root of the integration time, thousands of years would be needed for such a pulsar being detected by ET-D~\citep[see][for its sensitivity curve]{2011CQGra..28i4013H}.

\section{Conclusions and final remarks}\label{sec:5}

In this paper, we extend our previous studies ~\citep{2016arXiv160305975D,2016EPJC...76..481D} on the pulsar spindown in which we considered a combination of GW and magnetic energy dipole loss mechanisms. In particular, we explore in the present paper some consequences of an ellipticity generated by the magnetic dipole of the pulsar itself. It is well known that for magnetic fields of large strengths ($\sim10^{12}-10^{15}$~G), the equilibrium configuration of a neutron star can be distorted due to the magnetic tension.

Then, we firstly study the role of a time dependent $\epsilon_{B}$ on the calculation of the pulsars' braking indices. We argue that a time dependent $\epsilon_{B}$ stems naturally from the fact that such a quantity depends on the strength of the magnetic field and on the angle between the magnetic and rotation axes, which can well be time dependent.
This time dependent $\epsilon_{B}$ modifies the equation that relates $n$, $\eta$, etc. Instead of a multiplicative factor $(1-\eta)$ in equation \ref{neta}, we now have a factor of $(1+\eta)$. 

Secondly, we consider the role of the aforementioned deformation in the putative generation of GWs by the pulsars. 
In particular, we obtain useful equations, \ref{eek2} and \ref{ek2}, with which one can calculate $\epsilon_{B}$ and $\eta$
in terms of $I$, $M$, $R$, $\kappa$ and the observable quantities $P$ and $\dot P$.

From equation \ref{eek2}, we find that $\epsilon_{B}$
is extremely small ($ < 10^{-5}$) for the pulsars of Table \ref{ta1}, even for an unrealistic $\kappa \sim 10^3$. In addition, from equation \ref{ek2} one notices that 
$\eta < 10^{-8}$ for these very pulsars. Consequently, the amplitudes of the GWs for these pulsars are at best four orders of magnitude smaller than that obtained by assuming the spindown limit. Therefore, these results suggest that even advanced GW observatories will not be able to detect the pulsars of Table \ref{ta1}.

In a publication to appear elsewhere, we intend to extend the present study, in particular, that related to the GWs, for all pulsars with measured $P$ and $\dot P$, and discuss their putative detections.

Last but not least, the conclusions related to the detectability
of GWs are obviously dependent on the ellipticity generated by  the magnetic dipoles of the pulsars themselves. Whether there is some mechanism that could generate larger ellipticities, the prospects for the detection of the  pulsars of Table \ref{ta1}
would be less pessimistic. \\

J.C.N.A thanks FAPESP (2013/26258-4) and CNPq (308983/2013-0) for partial support. J.G.C. acknowledges the support of FAPESP (2013/15088-0 and 2013/26258-4).  C.A.C. acknowledges PNPD-CAPES for financial support. We thank the anonymous referee for valuable comments and suggestions.





\end{document}